\documentclass[aps,prc,superscriptaddress,twoside,twocolumn,nofootinbib,showpacs,floatfix]{revtex4-1}
\usepackage{amsmath,amssymb}
\usepackage{graphicx}
\usepackage{subfigure}
\usepackage[colorlinks]{hyperref}
\usepackage[usenames,dvipsnames]{color}
\pdfpagewidth=\paperwidth
\pdfpageheight=\paperheight
\allowdisplaybreaks
\usepackage{newtxtext, newtxmath}
\usepackage{braket,bm}
\usepackage{multirow}
\usepackage{enumitem}
\usepackage[vcentermath]{youngtab}

\begin{document}

\title{\boldmath $d^*(2380)$ and its partners in a diquark model}

\author{Pan-Pan Shi}
\affiliation{School of Nuclear Science and Technology, University of Chinese Academy of Sciences, Beijing 100049, China}

\author{Fei Huang}
\affiliation{School of Nuclear Science and Technology, University of Chinese Academy of Sciences, Beijing 100049, China}

\author{Wen-Ling Wang}
\affiliation{School of Physics and Nuclear Energy Engineering, Beihang University, Beijing 100191, China}

\date{\today}

\begin{abstract}
The purpose of the present study was to explore the possibility of accommodating the $d^*(2380)$ and its flavor SU(3) partners in a diquark model. Proposing that $d^*(2380)$ is composed of three vector diquarks, its mass is calculated by use of an effective Hamiltonian approach and its decay width is estimated by considering the effects of quark tunneling from one diquark to the others and the decays of the subsequent two-baryon bound state. Both the obtained mass and decay width of $d^*(2380)$ are in agreement with the experimental data, with the unexpected narrow decay width being naturally explained by the large tunneling suppression of a quark between a pair of diquarks. The masses and decay widths of the flavor SU(3) partners of $d^*(2380)$ are also predicated within the same diquark scenario.
\end{abstract}



\maketitle

\section{Introduction}

In recent years, the WASA-at-COSY Collaboration has reported the $d^*(2380)$ resonance with quantum numbers $I(J^P) = 0(3^+)$, mass $M \approx 2380$ MeV and width $\Gamma \approx 70$ MeV in double pionic fusion reactions $pn\to d\pi^0\pi^0$, $pn\to d\pi^+\pi^-$, and in other two pion production reactions e.g. $pn \to pn\pi^0\pi^0$, $pn\to pp\pi^-\pi^0$, $pd\to {^3\rm He}\,\pi^0\pi^0$, $pd \to {^3\rm He}\,\pi^+\pi^-$, $dd\to {^4\rm He}\,\pi^0\pi^0$ and $dd\to {^4\rm He}\,\pi^+\pi^-$ \cite{WASA2011,WASA2013,WASA2014-3,WASA2015,WASA2006,WASA2012,WASA2009} (See Ref.~\cite{Clement2016} for an experimental review). In a partial wave analysis of the $np$ scattering performed by the WASA-at-COSY Collaboration and the SAID Data Analysis Group, a resonance pole at $(2380 \pm 10) - i (40\pm 5)$ MeV in the ${^3D_3}-{^3G_3}$ coupled-channel waves has also been obtained after incorporating the newly observed analyzing power data of the polarized $\vec{n}p$ scattering \cite{WASA2014}.

Theoretically, the $d^*(2380)$ has so far attracted a lot of interests in hadron physics community \cite{Bashkanov:2013,Gal:2014zia,Huang:2013nba,Chen:2014vha,Huang:2014kja,Huang:2016,Dong:2016rva,Lu:2017uey,Dong:2017,Dong:2017geu,An:2016,Niskanen:2017}, with one of the main reasons being that it has an unexpected narrow decay width. As it is known, the $d^*(2380)$ is about $84$ MeV below the $\Delta\Delta$ threshold, but it is still much higher than the thresholds of the $\Delta N\pi$, $NN\pi \pi$ and $NN$ channels. Naively a rather wide decay width is expected since $d^*(2380)$ can decay to those channels via strong interactions. However, the experimentally observed value is only about $70$ MeV, which is even smaller than $1/3$ of the width of two $\Delta$'s. Such an unusual narrow decay width implies that the $d^*(2380)$ may have an unconventional structure involving new physical mechanisms.

In Refs.~\cite{Huang:2014kja,Huang:2016}, the $d^*(2380)$ is suggested to be a hexaquark dominated exotic state, as it has about $2/3$ hidden-color (CC) components in its configurations. Since the hidden-color configuration cannot decay directly into colorless hadrons at the lowest order, this picture automatically results in narrow decay width of $d^*(2380)$ \cite{Dong:2016rva,Dong:2017,Dong:2017geu}, a feature consistent with the experimental observation. Nevertheless, due cautions may still need to be taken in connection with the explanation of configuration structure of the bound $\Delta\Delta$-CC system. As pointed out in Ref.~\cite{Huang:2018}, the stability condition of a single baryon should always be satisfied in order to make meaningful discussions of the structure of a bound baryon-baryon system in quark model calculations.  

The purpose of the present study was to explore the possibility of accommodating the $d^*(2380)$ and its flavor SU(3) partners in a diquark model. The diquark model was first proposed by Gell-Mann in Ref.~\cite{GellMann:1964nj}, and has then been widely used to describe the structures of baryons and multiquark states in subsequent decades \cite{Anselmino:1992vq}. In particular, many recently observed $XYZ$ states have been proposed as multiquark states composed of diquark and anti-diquark \cite{Maiani:2004uc,Lebed:2017}. The most attractive feature of accommodating $d^*(2380)$ in a diquark model is that such a picture, if it works, will naturally produce a narrow decay width, a so far unintelligible feature awaiting reasonable theoretical explanations, since in a diquark model the decay of $d^*(2380)$ occurs in such a way that firstly the quarks tunnel from a diquark to the other two diquarks to form a $\Delta\Delta$ bound state and then the subsequent bound $\Delta\Delta$ decays, while the tunneling process highly suppresses the production probability and thus largely reduces the decay width.  

It is known that at small interquark separations, the force between two quarks arising from QCD is attractive in a color-antitriplet state, while it is repulsive in a color-sextet state. We thus confine ourselves to color-antitriplet diquarks only. We construct the diquarks so that the two quarks in each diquark are antisymmetric while any pair of diquarks are symmetric due to the Pauli principle. For $d^*(2380)$, since its quantum numbers are $I(J^P) = 0(3^+)$, we propose that it is composed of three vector diquarks. Each diquark belongs to a color antitriplet and has spin 1 and isospin 1. Two of the diquarks couple to form a color-triplet state with spin 2 and isospin 1, which, in turn couples to the third diquark to form a color-singlet state with spin 3 and isospin 0. The mass of $d^*(2380)$ is then calculated by use of an effective Hamiltonian approach motivated by QCD, with the model parameters fixed by known properties of heavy mesons and baryons. The decay width is roughly estimated by considering the process that a quark tunnels from one diquark to another to form a bound $\Delta\Delta$ state and then the subsequent bound $\Delta\Delta$ decays. It is found that both our obtained mass and decay width of $d^*(2380)$ are in agreement with the experimental data. Predictions are then made for the masses and decay widths of the flavor SU(3) partners of $d^*(2380)$ within the same diquark scenario.

The paper is organized as follows. In Sec.~\ref{Sec:Mass}, the effective Hamiltonian of the employed diquark model is introduced and the masses of $d^*(2380)$ and its flavor SU(3) partners are calculated. In Sec.~\ref{Sec:Width}, the widths of $d^*(2380)$ and its flavor SU(3) partners are estimated. Finally, a summary is given in Sec.~\ref{Sec:Summary}.

\section{The masses of $d^*(2380)$ and its flavor SU(3) partners} \label{Sec:Mass}

\subsection{The wave functions}

Due to the Pauli principle, two quarks in an orbital $S$-wave can stay in the following four possible configurations: $\left(0, {\textbf{6}}_f, {\textbf{6}}_c\right)$, $\left(1, \bar{\textbf{3}}_f, {\textbf{6}}_c\right)$, $\left(0, \bar{\textbf{3}}_f, \bar{\textbf{3}}_c\right)$, and $\left(1, {\textbf{6}}_f, \bar{\textbf{3}}_c\right)$. Here in each parenthesis, the first number denotes the spin of two quarks, the second and third symbols depict the irreducible representations of two-quark states in flavor and color spaces, respectively. The possible two-quark configurations and the corresponding Young diagrams in spin, flavor and color spaces are listed in Table~\ref{two-quark-GR}.

\begin{table}[tb]
\caption{\label{two-quark-GR} Possible two-quark configurations.}
\begin{tabular*}{\columnwidth}{@{\extracolsep\fill}lccc}
\hline\hline
                    & Spin           &   Flavor   &  Color         \\ 
\hline
\noalign{\vskip 4pt}
 $\left(0, {\textbf{6}}_f, {\textbf{6}}_c\right)$    &  $\yng(1,1)$  &  $\yng(2)$  &  $\yng(2)$    \\[9pt]
 $\left(1, \bar{\textbf{3}}_f, {\textbf{6}}_c\right)$    &  $\yng(2)$  &  $\yng(1,1)$  &  $\yng(2)$    \\[9pt]
 $\left(0, \bar{\textbf{3}}_f, \bar{\textbf{3}}_c\right)$    &  $\yng(1,1)$  &  $\yng(1,1)$  &  $\yng(1,1)$    \\[9pt]
 $\left(1, {\textbf{6}}_f, \bar{\textbf{3}}_c\right)$    &  $\yng(2)$  &  $\yng(2)$  &  $\yng(1,1)$    \\[9pt]
\hline\hline
\end{tabular*}
\end{table}

It is known that at small interquark separations, only in a color-antitriplet state the force between two quarks arising from QCD is attractive \cite{Maiani:2004vq}. And moreover, the spin of $d^*(2380)$ is 3, which requires that the constituent diquark should have spin 1. Therefore, we confine ourselves to the diquarks with configurations $\left(1, {\textbf{6}}_f, \bar{\textbf{3}}_c\right)$. The Pauli principle and the quantum numbers of $d^*(2380)$ require that any such two diquarks should couple to form a color ${\textbf{3}}_c$, spin 2 (or ${\textbf{5}}_s$) and flavor ${\textbf{15}}_f$ state, and finally, this two-diquark state couples to the third diquark to form a color singlet, spin 3 and flavor $\bar{\textbf{10}}_f$ sate. The wave function of three-diquark configurations in color, spin and flavor spaces are illustrated as follows:

\begin{enumerate}[label=(\roman*)]

\item Color wave function

$\left\{ \, \left[ ~ \yng(1,1) ~ \otimes ~ \yng(1,1) ~ \right]_{~ \scriptsize \yng(2,1,1)} ~ \otimes ~ \yng(1,1) ~ \right\}_{~ \scriptsize \yng(2,2,2)}$

\item Spin wave function

$\left\{ \, \left[ ~ \yng(2) ~ \otimes ~ \yng(2) ~ \right]_{~ \scriptsize \yng(4)} ~ \otimes ~ \yng(2) ~ \right\}_{~ \scriptsize\yng(6)}$

\item Flavor wave function

$\left\{ \, \left[ ~ \yng(2) ~ \otimes ~ \yng(2) ~ \right]_{~ \scriptsize \yng(3,1)} ~ \otimes ~ \yng(2) ~ \right\}_{~ \scriptsize\yng(3,3)}$

\end{enumerate}

\begin{figure}
\includegraphics[width=0.7\columnwidth]{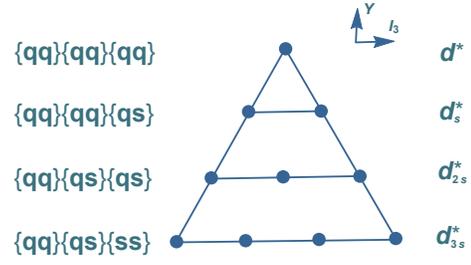}
\caption{Weight diagram of $\bar{\textbf{10}}_f$ of the flavor SU(3) group. The componential diquarks of each isospin multiplets are labeled at the left-hand side. The symbols $d^*_s$, $d^*_{2s}$ and $d^*_{3s}$ are introduced as the names of $d^*(2380)$'s partners with 1, 2, and 3 strange quarks, respectively, which have been marked at the right-hand side.}   
\label{fig:anti-decuplet} 
\end{figure}

The weight diagram of $\bar{\textbf{10}}_f$ of the flavor SU(3) group is plotted in Fig.~\ref{fig:anti-decuplet}, with the componential diquarks of each isospin multiplets are labeled at the left-hand side of this figure. There, $q$ represents $u$ or $d$ quark and $s$ the strange quark. The curly brackets indicate that the two quarks are symmetric in flavor space. For conciseness of the following parts of this paper, we have introduced the symbols $d^*_s$, $d^*_{2s}$ and $d^*_{3s}$ as the names of $d^*(2380)$'s partners with 1, 2 and 3 strange quarks, respectively, which have been marked on the right-hand side of Fig.~\ref{fig:anti-decuplet}.

The total wave functions of $d^*(2380)$ and its flavor SU(3) partners can be expressed explicitly as
\begin{align}
\Psi_{d^*} &= {\cal S} \! \left[ \left( \left\{q_1 q_2\right\}_{1,\bar{\textbf{3}}_c} \left\{q_3 q_4\right\}_{1,\bar{\textbf{3}}_c} \right)_{2,{\textbf{3}}_c} \left\{q_5 q_6\right\}_{1,\bar{\textbf{3}}_c} \right]_{3,{\textbf{0}}_c},  \label{eq:WF_d} \\[3pt]
\Psi_{d^*_s} &= {\cal S} \! \left[ \left( \left\{q_1 q_2\right\}_{1,\bar{\textbf{3}}_c} \left\{q_3 q_4\right\}_{1,\bar{\textbf{3}}_c} \right)_{2,{\textbf{3}}_c} \left\{q_5 s\right\}_{1,\bar{\textbf{3}}_c} \right]_{3,{\textbf{0}}_c},  \label{eq:WF_ds} \\[3pt]
\Psi_{d^*_{2s}} &= {\cal S} \! \left[ \left( \left\{q_1 q_2\right\}_{1,\bar{\textbf{3}}_c} \left\{q_3 s\right\}_{1,\bar{\textbf{3}}_c} \right)_{2,{\textbf{3}}_c} \left\{q_4 s\right\}_{1,\bar{\textbf{3}}_c} \right]_{3,{\textbf{0}}_c},  \label{eq:WF_d2s} \\[3pt]
\Psi_{d^*_{3s}} &= {\cal S} \! \left[ \left( \left\{q_1 q_2\right\}_{1,\bar{\textbf{3}}_c} \left\{q_3 s\right\}_{1,\bar{\textbf{3}}_c} \right)_{2,{\textbf{3}}_c} \left\{s s\right\}_{1,\bar{\textbf{3}}_c} \right]_{3,{\textbf{0}}_c},  \label{eq:WF_d3s}
\end{align}
where the first and second subscripts of each diquark indicate the spin and irreducible representation of color, respectively. The symbol ${\cal S}$ is the symmetrizer for three diquarks. Note that by construction, the diquarks are already symmetric in spin space and antisymmetric in color space, and therefore they should be antisymmetric in flavor space.

\subsection{The mass formulas}

In literature, the phenomenological Hamiltonian of a diquark model is usually parametrized as a sum of the diquark masses and an effective potential, with the potential being composed of spin-spin interaction in heavy quark systems \cite{Maiani:2004vq,Zhu:2015bba} and color-electric and color-spin terms, inspired by one-gluon exchange potential and instanton-induced interaction, in light quark systems \cite{Kim:2016dfq}. Following these references, we express the Hamiltonian employed in the present work as
\begin{equation}  \label{eq:Hamiltonian}
H = \sum_n M_n + 2\sum_{i>j} \left[ \alpha_{ij} \left( {\bm \lambda}_i \cdot {\bm \lambda}_j {\bm S}_i \cdot {\bm S}_j \right) + \frac{\beta}{m_im_j} \left( {\bm \lambda}_i \cdot {\bm \lambda}_j \right) \right],
\end{equation}
where the coefficient $\alpha_{ij}$ and the masses $m_i$ and $m_j$ depend on the flavor of the constituents $i$ and $j$, the coefficient $\beta$ is flavor independent, and $M_n$ is the effective mass of the $n$-th constituent which includes also those effects not accounted for by the above-mentioned interactions.

By use of the wave functions given in Eqs.~(\ref{eq:WF_d})-(\ref{eq:WF_d3s}), one gets the following matrix elements for $d^*(2380)$ and its partners,
\begin{align}
& \Braket{{\bm \lambda}_i \cdot {\bm \lambda}_j} = \left\{ \begin{array}{lll} -8/3, && \left(i, j {\rm ~ in ~ the ~ same ~ diquark}\right) \\[2pt] -2/3, && \left({\rm others}\right) \end{array} \right.  \\[3pt]
& \Braket{{\bm \lambda}_i \cdot {\bm \lambda}_j {\bm S}_i \cdot {\bm S}_j} = \frac{1}{4}\Braket{{\bm \lambda}_i \cdot {\bm \lambda}_j}.
\end{align}
The masses of $d^*(2380)$ and its partners can then be expressed as:
\begin{align}
M_{d^*} =&\, 3M_{qq} - 8\alpha_{qq} - \frac{32\beta}{m_q^2},  \label{eq:M_d} \\[3pt]
M_{d^*_s} =&\, 2M_{qq} + M_{qs} -\frac{16}{3}\alpha_{qq} -\frac{8}{3}\alpha_{qs} - \frac{64\beta}{3m_q^2} \nonumber \\
& - \frac{32\beta}{3m_qm_s},  \label{eq:M_ds} \\[3pt]
M_{d^*_{2s}} =&\, M_{qq} + 2M_{qs} -3\alpha_{qq} -\frac{14}{3}\alpha_{qs} -\frac{1}{3}\alpha_{ss}  \nonumber \\
& -\frac{12\beta}{m_q^2} -\frac{56\beta}{3m_qm_s} -\frac{4\beta}{3m_s^2},  \label{eq:M_d2s} \\[3pt]
M_{d^*_{3s}} =&\, M_{qq} + M_{qs} + M_{ss} -2\alpha_{qq} -4\alpha_{qs} -2\alpha_{ss}  \nonumber \\
& -\frac{8\beta}{m_q^2} -\frac{16\beta}{m_qm_s} -\frac{8\beta}{m_s^2}.  \label{eq:M_d3s}
\end{align}
Following Jaffe \cite{Jaffe2005}, the diquark mass $M_{ss}$ is related to $M_{qq}$ and $M_{qs}$ by
\begin{equation}
M_{ss} = 2M_{qs} - M_{qq}.   \label{eq:M_ss}
\end{equation}
The diquark masses $M_{qq}$, $M_{qs}$ and the coefficients $\alpha_{qq}$, $\alpha_{qs}$, $\alpha_{ss}$, $\beta$ are model parameters to be fixed in the next subsection.

\subsection{The model parameters}

Following Jaffe \cite{Jaffe2005}, the masses of the heavy baryons $\Lambda_c$, $\Xi_c$, $\Omega_c^0$, $\Sigma_c$, $\Xi'_c$, $\Sigma_c^*$ and $\Xi_c^*$ which are attributed to diquark-quark configurations, and the masses of the heavy mesons $D_s$ and $D_s^*$ which are treated as quark-antiquark configurations, are used to fix the parameters of the diquark masses and the coefficients required in Eqs.~(\ref{eq:M_d})-(\ref{eq:M_d3s}). Specifically, presuming that $\Lambda_c$, $\Sigma_c$ and $\Sigma_c^*$ are composed of $\left([qq]c\right)^{J=1/2}$, $\left(\{qq\}c\right)^{J=1/2}$ and $\left(\{qq\}c\right)^{J=3/2}$, respectively, with $[~]$ representing antisymmetric and $\{~\}$ symmetric in flavor space for two quarks in a diquark, the following expressions for the coefficients $\alpha_{qq}$, $\beta$ and the diquark mass $M_{qq}$ can be obtained after applying the Hamiltonian of Eq.~(\ref{eq:Hamiltonian}) to the heavy baryons $\Lambda_c$, $\Sigma_c$, $\Sigma_c^*$ and to the heavy mesons $D_s$, $D_s^*$:
\begin{align}
\alpha_{qq} &= -\frac{m_{\Sigma_c}+2m_{\Sigma_c^*}-3m_{\Lambda_c}}{16},  \label{eq:alpha_qq}  \\[3pt]
\beta & = \frac{3m_sm_c}{128}\left(4m_s+4m_c-m_{D_s}-3m_{D_s^*}\right),   \label{eq:beta}  \\[3pt]
M_{qq} & = m_{\Lambda_c}-m_c+\frac{16\beta}{3m_q^2m_c}\left(m_c+2m_q\right)-4\alpha_{qq}.  \label{eq:M_qq}
\end{align}
Similarly, supposing that $\Xi_c$, $\Xi_c'$ and $\Xi_c^*$ are composed of $\left([qs]c\right)^{J=1/2}$, $\left(\{qs\}c\right)^{J=1/2}$ and $\left(\{qs\}c\right)^{J=3/2}$, respectively, the following expressions for the coefficient $\alpha_{qs}$ and the diquark mass $M_{qs}$ can be obtained after applying the Hamiltonian of Eq.~(\ref{eq:Hamiltonian}) to these three heavy baryons:
\begin{align}
\alpha_{qs} &= -\frac{m_{\Xi'_c}+2m_{\Xi_c^*}-3m_{\Xi_c}}{16},    \label{eq:alpha_qs}   \\[3pt]
M_{qs}  & = m_{\Xi_c}-m_c + \frac{16\beta}{3m_qm_sm_c} \left(m_c+m_s+m_q\right)-4\alpha_{qs}. \label{mqs}
\end{align}
Finally, postulating that $\Omega_c^0$ is composed of $\left(\{ss\}c\right)^{J=3/2}$, the following expression of the coefficient $\alpha_{ss}$ can be obtained after applying the Hamiltonian of Eq.~(\ref{eq:Hamiltonian}) to the $\Omega_c^0$ baryon:
\begin{equation}
\alpha_{ss} = -\frac{1}{4}\left[M_{ss}+m_c-\frac{16\beta}{3m_s^2m_c} \left(m_c+2m_s\right)-m_{\Omega_c^0}\right],\label{ass}
\end{equation}
where $M_{ss}$ is related to $M_{qq}$ and $M_{qs}$ by Eq.~(\ref{eq:M_ss}).

\begin{table}[tb]
\caption{\label{parameter} Model parameters. The diquark masses $M_{qq}$, $M_{qs}$ and the coefficients  $\alpha_{qq}$, $\alpha_{qs}$, $\alpha_{ss}$ are in MeV. The coefficient $\beta$ is in fm$^{-3}$. }
\renewcommand{\arraystretch}{1.2}
\begin{tabular*}{\columnwidth}{@{\extracolsep\fill}cccccc}
\hline \hline
  $M_{qq}$   &   $M_{qs}$  &  $\alpha_{qq}$  &  $\alpha_{qs}$   &  $\alpha_{ss}$   &  $\beta$   \\
   $1032$     &   $1103$     &   $-39.5$         &   $-29.1$           &  $-3.3$            &  $0.41$    \\
\hline\hline
\end{tabular*}
\end{table}

In the present work, the masses of $u$, $d$, $s$ quarks are taken to be $m_{u(d)}=313$ MeV and $m_s=470$ MeV, values used in Ref.~\cite{Huang:2018} which gives a satisfactory description of the masses of octet and decuplet baryon ground states, the binding energy of deuteron, and the $NN$ scattering phase shifts up to a total angular momentum $J=6$ in a rather consistent manner. For the mass of $c$ quark, we use $m_c=1650$ MeV, an averaged value of $1600\sim 1700$ MeV which is commonly used in literature. By implementing the experimental values of the masses of $\Lambda_c$, $\Sigma_c$, $\Sigma_c^*$, $\Xi_c$, $\Xi'_c$, $\Xi_c^*$, $\Omega_c^0$, $D_s$ and $D_s^*$ into Eqs.~(\ref{eq:alpha_qq})-(\ref{ass}), one gets the values of the model parameters $M_{qq}$, $M_{qs}$, $\alpha_{qq}$, $\alpha_{qs}$, $\alpha_{ss}$ and $\beta$, which are listed in Table~\ref{parameter}.  Here we mention that the diquark masses $M_{qq}$ and $M_{qs}$ include all the effects that are not considered in Eq.~(\ref{eq:Hamiltonian}), e.g. the kinematic energies and the confinement {\it et al.}. Therefore their values are bigger than the masses of two constituent quarks, although the interactions between them are attractive.

\subsection{Numerical results}

With the values of model parameters listed in Table~\ref{parameter}, the masses of $d^*(2380)$ and its flavor SU(3) partners can be obtained by use of Eqs.~(\ref{eq:M_d})-(\ref{eq:M_d3s}). The numerical results are listed in Table~\ref{mass}. One sees that the mass of $d^*(2380)$, $2383$ MeV, is very close to the experimental value, $2380$ MeV. In the third column of this table, we list the possible baryon-baryon channels that $d^*(2380)$ and its flavor SU(3) partners can transform into when quarks in a diquark tunnel into another two diquarks. The corresponding mass thresholds of these baryon-baryon channels are listed in the fourth column of this table. In the last column, we list the differences of columns 2 and 4, which are binding energies of the corresponding baryon-baryon states formed after the tunneling processes. One sees that the calculated mass of $d^*$ is $81$ MeV lower than the threshold of $\Delta\Delta$, while the predicated masses of $d^*_{s}$, $d^*_{2s}$ and $d^*_{3s}$ are $76$ MeV, $73$ MeV and $107$ MeV lower than the thresholds of $\Delta \Sigma^*$, $\Delta \Xi^*$ and $\Delta \Omega$, respectively. 

\begin{table}[tb]
\caption{\label{mass} Masses of $d^*(2380)$ and its flavor SU(3) partners. The masses $M$, the corresponding baryon-baryon thresholds $M_{\rm thr.}$, and their differences $M_{\rm thr.}-M$ are in MeV.}
\renewcommand{\arraystretch}{1.2}
\begin{tabular*}{\columnwidth}{@{\extracolsep\fill}lcccc}
\hline\hline
                      &  $M$       &  channel                  & $M_{\rm thr.}$  &   $M_{\rm thr.}-M$         \\
\hline
  $d^*$           &  $2383$   &  $\Delta \Delta$       &  $2464$     &  $81$      \\
  $d_s^*$        &  $2541$    &  $\Delta \Sigma^*$  &  $2617$      &  $76$      \\
  $d_{2s}^*$    &  $2689$    &  $\Delta \Xi^*$        &  $2762$     &  $73$       \\
  $d_{3s}^*$    &  $2797$   &  $\Delta \Omega$    &  $2904$      &  $107$     \\[2pt]
\hline\hline
\end{tabular*}
\end{table}

\section{The decay widths of $d^*(2380)$ and its flavor SU(3) partners}  \label{Sec:Width}

The experimental decay width of $d^*(2380)$, $\Gamma_{d^*}\sim 70$ MeV, is unexpectedly narrow, as $\Delta$ is rather broad ($\Gamma_\Delta\sim 117$ MeV) and the mass of $d^*(2380)$ is above the thresholds of $\Delta N\pi$, $NN\pi\pi$ and $NN$ channels even it is about $84$ MeV below the $\Delta\Delta$ threshold. Although such an unexpected narrow decay width might be a challenge for hadron physicists when explain the $d^*(2380)$ in a traditional picture like a $\Delta\Delta$ bound state, in the diquark scenario as proposed in the present work, this narrow decay width can be straightforwardly interpreted, as in this picture the decay occurs in such a way that the quarks in a diquark tunnel into another two diquarks to form a bound $\Delta\Delta$ sate and then the subsequent bound $\Delta\Delta$ decays, while the tunneling process is highly suppressing against the decay rates. 

In the present work we would perform a rough estimate of the decay widths of $d^*(2380)$ and its flavor SU(3) partners instead of making an accurate calculation.

Using the leading semiclassical approximation, the tunneling amplitudes read \cite{Landau1977,Maiani2018}
\begin{align}
{\cal A}(r) \sim e^{-r\sqrt{2m{\Delta E}}},  \label{eq:amplitude_tunneling}
\end{align}
with $\Delta E$ being the depth of the potential barrier that a quark in a diquark needs to tunnel through, and $m$ being the mass of tunneling quark. In our case, $\Delta E$ can be approximated by
\begin{align}
\Delta E \approx \frac{1}{3} \left( M_{\rm thr.}-M \right),
\end{align}
with $M$ being the mass of a three-diquark state and $M_{\rm thr.}$ the mass threshold of the corresponding baryon-baryon state that a three-diquark state can transform into after the tunneling process, as listed in Table~\ref{mass}. 

From Eq.~(\ref{eq:amplitude_tunneling}), one gets the tunneling probability
\begin{align}
{\cal P}^2 \sim \frac{\int_{r_0/2}^\infty \left|{\cal A}(r)\right|^2 {\rm d}r }{\int_0^\infty \left|{\cal A}(r)\right|^2 {\rm d}r} = e^{-r_0\sqrt{2m{\Delta E}}},
\end{align}
with $r_0$ being the distance of two diquarks.  Then the decay widths of $d^*$ and its flavor SU(3) partners can be approximated by
\begin{align}
\Gamma \approx {\cal P}^2 \Gamma_{BB'} \approx e^{- r_0\sqrt{2m\left( M_{\rm thr.}-M \right)/3} } \left(\Gamma_{B} + \Gamma_{B'} \right),    \label{eq:width_dst}
\end{align}
with $\Gamma_{B(B')}$ being the decay width of a bound baryon $B(B')$.

\begin{table}[tb]
\caption{\label{radius} Decay widths (in MeV) of $d^*(2380)$ and its flavor SU(3) partners at selected values for the distances (in fm) of two diquarks. $\Gamma_{BB'}$ is the decay width of a bound $BB'$ state.}
\renewcommand{\arraystretch}{1.2}
\begin{tabular*}{\columnwidth}{@{\extracolsep\fill}lrrrrrrr}
\hline\hline
                 & \multirow{2}{*}{$\Gamma_{BB'}$}    &  \multicolumn{6}{c}{$r_0$}     \\   \cline{3-8}                                            
                 &    &   $0.9$  &  $1.1$  & $1.3$ & $1.5$ & $1.7$   & $1.9$   \\                                               
\hline
   $d^*$             & 168.2       & 93.0    & 81.5      & 71.4     & 62.6      & 54.9      & 48.1     \\
   $d^*_s$          & 107.9       & 60.7     & 53.5      & 47.1     & 41.4      & 36.5    & 32.1         \\
   $d^*_{2s}$      & 88.7        & 50.2    & 44.2      & 39.0   & 34.3        & 30.3   & 26.7      \\
   $d^*_{3s}$      & 66.6        & 33.7     & 29.0     & 24.9     & 21.4      & 18.4     & 15.8    \\[2pt]
\hline\hline
\end{tabular*}
\end{table}

The width of a bound $\Delta$ can be parametrized as \cite{Pilkuhn67}
\begin{align}
\Gamma_\Delta = \gamma_{N\pi}  \frac{R^2q^2_{N\pi}}{1+R^2q^2_{N\pi}} q_{N\pi},  \label{eq:width_Delta}
\end{align}
with $q_{N\pi}$ being the magnitude of the momentum of $\pi$ or $N$ in $\Delta$ rest frame, $R=6.3$ (GeV/c)$^{-1}$, and $\gamma_{N\pi}=0.76$ which ensures that for a free $\Delta$ the calculated decay width equals the experimental value. For $\Delta$'s decuplet partners, $\Sigma^*$ and $\Xi^*$, we assume that their widths can be parametrized in a similar way as $\Delta$, i.e.
\begin{align}
\Gamma_{\Sigma^*} &= \gamma_{\Lambda\pi}  \frac{R^2q^2_{\Lambda\pi}}{1+R^2q^2_{\Lambda\pi}} q_{\Lambda\pi} + \gamma_{\Sigma\pi}  \frac{R^2q^2_{\Sigma\pi}}{1+R^2q^2_{\Sigma\pi}} q_{\Sigma\pi}, \label{eq:width_Sigst}  \\[3pt]
\Gamma_{\Xi^*} &= \gamma_{\Xi\pi}  \frac{R^2q^2_{\Xi\pi}}{1+R^2q^2_{\Xi\pi}} q_{\Xi\pi},  \label{eq:width_Xist}
\end{align}
with $\gamma_{\Lambda\pi}=0.24$, $\gamma_{\Sigma\pi}=0.09$ and $\gamma_{\Xi\pi}=0.13$ which ensures that for free cases the calculated total or partial decay withs are exactly the same as the experimental values. Note that the bound $\Omega$ baryon is rather stable as the free $\Omega$ has nearly $0$ MeV decay width.

By using Eq.~(\ref{eq:width_dst}) together with Eqs.~(\ref{eq:width_Delta})-(\ref{eq:width_Xist}), one gets the approximated values of the decay widths for $d^*$ and its flavor SU(3) partners as a function of the distance of two diquarks, and the numerical results are listed in Table~\ref{radius}. One sees that when the distance of two diquarks $r_0 \sim 1.3$ fm, the estimated width of $d^*(2380)$ will be close to its experimental value. At this distance, the predicated widths of $d^*_{s}$, $d^*_{2s}$ and $d^*_{3s}$ are abound $47$ MeV, $39$ MeV and $25$ MeV, respectively. Of course this is just a very preliminary and rough estimation. A rigorous calculation is beyond the scope of the present work, but is planned for the near future.

\section{Summary}   \label{Sec:Summary}

The unexpected narrow decay width of $d^*(2380)$ makes it a challenge for hadron physicists to interpret this particle with a conventional picture. The purpose of the present work was to explore the possibilities to accommodate the $d^*(2380)$ and its flavor SU(3) partners in a diquark scenario. 

We propose that $d^*(2380)$ and its flavor SU(3) partners are composed of three vector diquarks, and construct their color, spin, flavor wave functions and further the total wave functions according to Pauli principle. The masses of $d^*(2380)$ and its flavor SU(3) partners are calculated by use of an effective Hamiltonian with QCD-inspired interactions. The decay widths are estimated by a two-step mechanism, i.e. firstly a quark tunnels from one diquark to another resulting in a baryon-baryon bound state and then the subsequent bound baryon-baryon state decays. The former highly suppresses the decay rates and leads to a natural explanation of the narrow width of $d^*(2380)$.

Both the calculated mass and the estimated decay width of $d^*(2380)$ are in agreement with the data, not excluding the possibility of assigning a diquark picture to it. The masses and widths for $d^*$'s flavor SU(3) partners obtained in a similar way as $d^*(2380)$ serve as preliminary predictions for further theoretical and experimental investigations.

\begin{acknowledgments}
One of the authors, Pan-Pan Shi, is grateful to Feng-Kun Guo and Chao-Feng Liu for their useful discussions. This work is partially supported by the National Natural Science Foundation of China under Grant No.~11475181 and No.~11635009, the Youth Innovation Promotion Association of CAS under Grant No.~2015358, and the Key Research Program of Frontier Sciences of CAS under Grant No. Y7292610K1.
\end{acknowledgments}

\end{document}